\documentclass[doublecol,linenumbers]{epl2} 

\usepackage{amsmath}

\title{Imitation-Induced Criticality: Network Reciprocity  and Psychological Reward}
\shorttitle{social benefit of cooperation} 

\author{Korosh Mahmoodi,  Paolo Grigolini}
\shortauthor{}

\institute{Center for Nonlinear Science, University of North Texas, P.O. Box 311427,
Denton, Texas 76203-1427, USA\\ }

\abstract{The nodes of a regular two-dimensional lattice play a game based on the joint action of two distinct levels. At the first step of the game,  using a random prescription half players are assigned the cooperation and half the defection state. At the bottom level the strategy choice is done on the mere basis of imitation according to the \emph{homo imitans} principle, generating a form of collective intelligence that makes the system sensitive to  
 the criteria determining the strategy choice adopted at the 
 top level.   
 The units of the top level, in fact,  play the prisoner's dilemma game and are allowed to  update their strategy 
either by selecting the strategy of the most successful nearest neighbor, \emph{success model}, or merely on the basis of the criterion of the best financial benefit, \emph{selfishness model}. The intelligence emerging from imitation-induced criticality  leads in the former case to the extinction of defection and in the latter case to the extinction of cooperation. 
The former case is interpreted as a form of network reciprocity enhanced by the imitation-induced criticality and contributing to the evolution towards cooperation. We perturb the selfishness model with a form of morality pressure, exerted by a psychological reward $\lambda$ for cooperation, to establish the sensitivity of collective intelligence to morality.  We find that when $\lambda$ gets a crucial value $\lambda_c$, exceeding the temptation to cheat,
the system makes a transition from the supercritical defection state  to the critical regime, with the warning that an excess of morality and religion pressure may annihilate the criticality-induced resilience of the system.}

\begin{document}

\maketitle

\section{Introduction}

The unification of behavioral sciences is an attractive and challenging enterprise that would be impossible without using game theory \cite{1}. The recent book by Gintis \cite{1}  aims at the unification of Behavioral Sciences, namely, at the ambitious purpose of unifying biology, psychology, economics, anthropology and political science, stressing however that game theory alone is not enough to realize this important goal. Game theory is a theoretical attempt at explaining why the selfish action of single individuals is compatible with the emergence of altruism and cooperation. Nowak and May \cite{4} have attracted the attention of an increasing number of researchers on the network reciprocity, a special condition where spatial structure favors the emergence of altruism, see also \cite{2}, in spite of its nodes playing the prisoner's dilemma game \cite{3}, which is expected to favor defection. Actually, since the patch owner, either a cooperator $C$ or a defector $D$ is replaced by the neighbor with the largest payoff \cite{4}, this is equivalent to updating the strategy of each player adopting that of the most successful neighbor, and a cluster of cooperators has the effect of protecting the units of the cluster from the exploitation of the defectors. 

As illustrated in the recent review paper of Wang \emph{et al} \cite{5}, the research work in this field is now focusing on the topology of networks and especially on the emerging field of multilayers networks to explain the pattern formation with production of clusters protecting cooperators from the exploitation of the defectors so as to favor their survival of cooperators.  
This Letter adopts a multilayers perspective,  by using however \emph{dynamical} rather than merely topological arguments,  in such a way as to be as close as possible to the project of unification of behavioral sciences recently proposed by Grigolini \emph{et al} \cite{6,7}, as an attempt at addressing the challenge of Gintis \cite{1}. Furthermore the multiple layers do not necessarily correspond to different nodes, but in this Letter refer to different levels of human behavior, the social, the financial  and the spiritual. 
The bottom layer of this Letter is based on the observation that the individuals of a network playing the prisoner's dilemma game are the units of a human society and are expected to 
be strongly influenced by imitation \cite{8,9,10}. The individuals of this network make a choice between the cooperation state $C$ and the defection state $D$, without any form of cognition.  This choice is not determined by the wish of maximizing the personal benefit through imitation of the most successful neighbor \cite{2} or by the greedy choice of an immediate payoff, but imitation is as blind as the bird tendency to select their flying direction on the basis of the flying directions of the neighbor birds \cite{11}, with no consideration of the personal benefit,   either direct or indirect.  The imitation strength is a control parameter hereby denoted by the symbol $K_1$. A critical value of $K_1$, called $K_{1c}$,  exists making it possible for the swarm to fly as a whole. Although the action of the single individuals of the network does not require any form of cognition,  as stressed by the authors of Ref. \cite{12,13}, criticality  generates a form of collective intelligence.  This collective intelligence  is characterized not only by the criticality-induced 
long-range correlation but also, and especially, by temporal complexity \cite{14,15,16,17}, a condition making the complex network flexible and resilient.  It is important to stress that the supercritical 
condition is characterized by fluctuations around 
a non vanishing mean field as random as the fluctuations around the vanishing mean field of the subcritical regime, thereby lacking the flexibility and resilience of criticality. 

At the top layer the units play the prisoner's dilemma game \cite{3} and exert an influence on the bottom level choosing their strategy according to either the success or selfishness criterion. 

The \emph{success model} is realized as follows. For most of time steps the choice of strategy is determined by the bottom level, and from time to time the units are allowed to select their strategy adopting that of their most successful nearest neighbor, as suggested by the pioneer work of Ref. \cite{4}. 
 The success model yields the impressive effect of annihilating the emergence of the branch with the majority of defectors when the system adopting this choice is made intelligent by imitation-induced criticality. 
We are convinced that this effect affords a solid explanation of the 
evolution towards cooperation, this being a very important result of this Letter.

We compare the \emph{success model}  to the \emph{selfishness model}, where
the player does not adopt the strategy of the most successful nearest neighbor, but she makes her choice only on the basis of her personal benefit. She evaluates the financial benefit derived from the defection choice and the financial benefit that she would get from the cooperation choice, giving larger weight to the maximal profit. The benefit of a given choice is done assuming that the unit in action can play with equal probability with all her neighbors. The selection of the convenient strategy is weighted with a second control parameter $K_2$, with the ratio $\rho_K = K_2/K_1$ establishing if the link with the top layer is stronger, $\rho_K> 1$,  or weaker, $\rho_K< 1$, than the link with the bottom layer. 
Switching on the interaction with the top layer has  the effect of leading to the extinction of cooperation with a big loss for society, even if, as we shall see hereby, $K_1 > 0$ yields  imitation-induced clusters of cooperators  with  financial benefit for society, this being the reason why the \emph{success model}, for $K \gg K_{1c}$ leads to the extinction of defectors.   
The strategy choice determined by the criterion of maximal personal benefit, rather than by the choice of the strategy of the most successful nearest neighbor \cite{4}, on the contrary, yields the extinction of cooperation, with an even very small value of $\rho_K$, as an effect of 
imitation-induced intelligence.  

 To complete the illustration of the role of criticality-induced intelligence we study the influence of morality on the dynamics of the \emph{selfishness model}  showing that as an effect of imitation-induced intelligence the system becomes so sensitive to morality  as to make  a psychological reward moderately  exceeding the temptation to cheat,  robust enough as to prevent the collapse of the social system into the sub-criticality disorder. In principle the influence of morality on the system should be established by the interaction of the network with an 
additional layer. For simplicity's sake, we modify the conventional 
prisoner's dilemma game \cite{1,18} through the introduction of the psychological reward for the choice of cooperation. The strength of psychological,  called $\lambda$, affords a simplified way to describe 
the influence that an additional layer, concerning morality and religion, may have on the selfishness model. We find that, when the bottom layer operates at criticality,  
a crucial value $\lambda_c$ exists with the effect of preventing  the extinction of cooperators and of recovering the criticality-induced temporal complexity that is essential for the healthy behavior of the social system. We call this \emph{Asbiyyah} effect \cite{19}, this arabic world  meaning \emph{group feeling}, namely the natural tendency of human beings to cooperate. This natural disposition is  enhanced by religion and it has the eventual effect of increasing the social prosperity, but, in accordance with the observation of Ahmed \cite{20}, we find that $\lambda > \lambda_c$, the super-asbiyyan condition,  may be as bad as the lack of social cohesion. 

\section{Game theory}

As well known \cite{3}, game theory rests on the crucial inequality
\begin{equation}
T > R > P > S ,
\end{equation}
where $R$ denotes the reward that a cooperator gets when playing with another cooperator.  The parameter $T > R$ is the payoff of a defector: it is larger than $R$ thereby $t = T-R$ is a measure of the temptation to cheat. $S$ is the payoff of  a  cooperator playing with a defector, and $P < R$ is the payoff of a defector playing with another defector. The condition 
\begin{equation}
2 R > T + S
\end{equation} 
indicates that the community gets a larger benefit from the play between two cooperators than from the play between a cooperator and a defector. Of course, also the play between two defectors with $2R > 2P$ is less beneficial to the community than the play between two cooperators.
We adopt the choice made by Gintis \cite{1,18} and we set $R = 1$, $P = 0$, $T = 1 + t$ and $S = - s$. To study the influence of morality on the selfishness model we introduce the psychological reward $\lambda$, setting\begin{equation}
R = 1 + \lambda,
\end{equation}
\begin{equation}
T = 1 + t,
\end{equation}
\begin{equation}
P = - \lambda
\end{equation}
and
\begin{equation}
S = -s.
\end{equation}
In this Letter we adopt always but in Fig. \ref{NEW} the choice
\begin{equation}
s = 3
\end{equation}
and
\begin{equation}
t = 2. 
\end{equation}

Note that we denote by $N$ the total number of players. They are the nodes of a two-dimensional regular network 
of size $32$. Therefore throughout the whole Letter we adopt $N = 32X32$. We use the symbol $L$ to denote the number of time steps, ranging from $L = 10^4$ to $L = 10^6$. 
 \section{Decision-Making, Success, Selfishness and Influence of  Morality Model }
To establish an interaction between the bottom and the top level we adopt a natural extension of the \emph{Decision Making Model} (DMM) of Ref. \cite{8}. The transition rate from the cooperator to the defector state,  $g_{12}$, is given by
\begin{equation}
g_{12} =  g_0 exp \left[ - K_1 \left(\frac{M_1 - M_2}{M}\right) - K_2 \left(\frac{\Pi_C - \Pi_D}{|\Pi_C| + |\Pi_D|}\right) \right]
\end{equation}
and  the transition rate from the defector to the cooperator state, $g_{21}$, is given by
\begin{equation}
g_{21} = g_0 exp \left[  K_1 \left(\frac{M_1 - M_2}{M}\right)  + K_2 \left(\frac{\Pi_C - \Pi_D}{|\Pi_C| + |\Pi_D|}\right) \right].
\end{equation}
The meaning of this prescription is as follows. The parameter $1/g_0$ defines the time scale and we set $g_0 = 0.1$ throughout.  We consider $N$ units. Each unit has $M$ neighbors (four in the case of the regular two-dimensional lattice used in this Letter). The cooperation state corresponds to the state $|1>$ and the defector state 
 to the state $|2>$.  In the case $K_2 = 0$, this is the ordinary DMM of Ref. \cite{8}.  If the unit is in the cooperator state, $|1>$, and the majority of its neighbors 
are in the same state, then the transition rate becomes smaller and the units sojourns in the cooperation state for a longer time. If the majority of its neighbors
is in the defector state, then the unit that has to make a decision sojourns in the cooperator state for a shorter time. Analog prescription is used if the unit is in the defector state.
Note that at $K = K_{1c}$ the units move from a dynamical condition where they are virtually independent the ones from the others to a condition
where global order emerges. In the case of two-dimensional regular lattice of this article, ${K_1}_{c} \approx 1.65$.

\begin{figure}
\onefigure[width= 0.4 \textwidth]{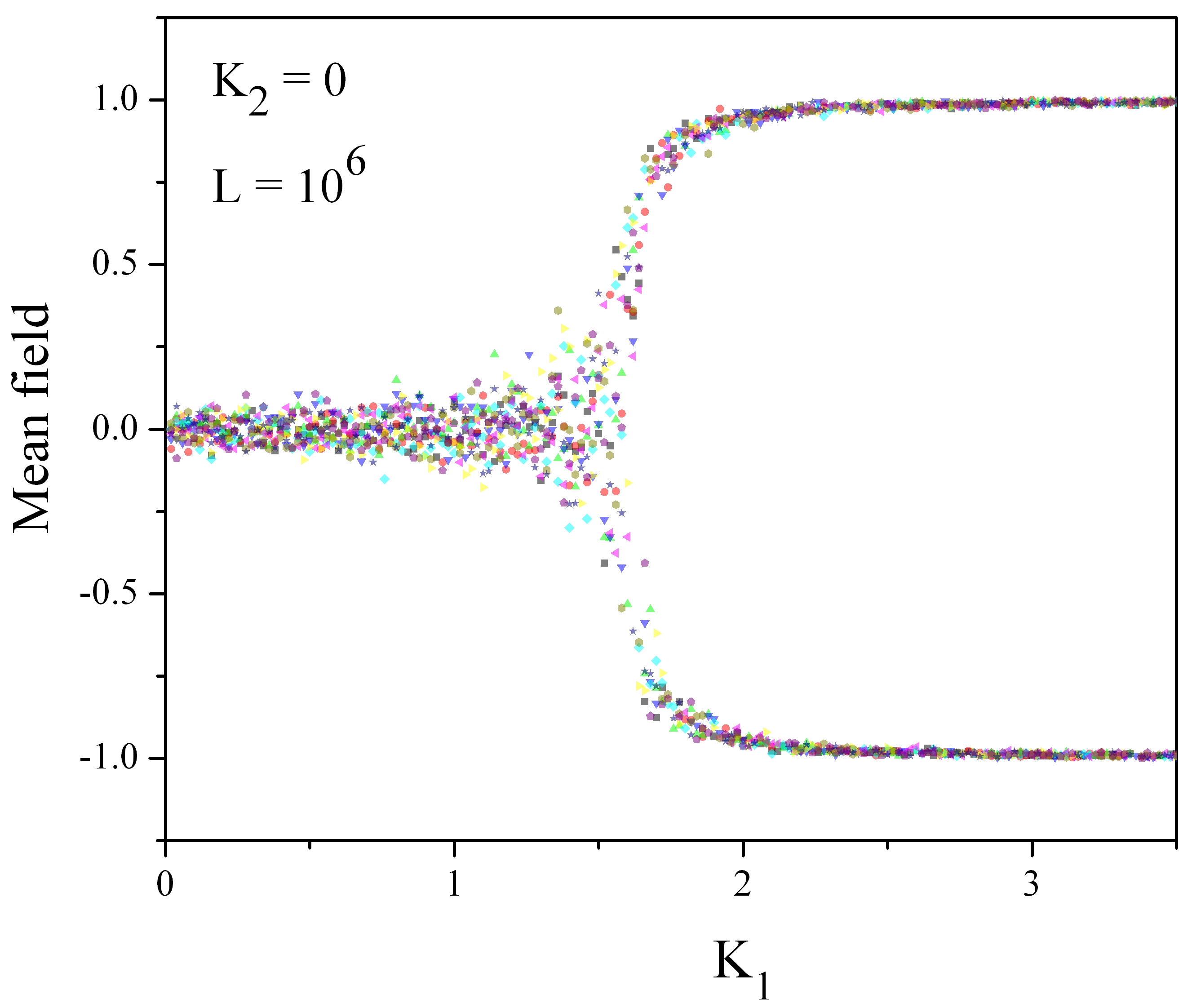} 
\caption{Mean field of the bottom level when $K_2 = 0$.   We make 10 realizations.}
\label{FIG1}
\end{figure}

Fig. \ref{FIG1} illustrates the second-order phase transition generated by the DMM, namely, the model of this Letter when $K_2 = 0$. This corresponds to the condition where no bias exists for the choice of either cooperation or defection. In the initial condition the $N$ units of the network are randomly assigned to either the cooperation or the defection state with the same probability, making zero  the mean field 
of the network
\begin{equation}
\left<\xi \right> \equiv \frac{\sum_{i}^N \xi_i}{N},
\end{equation}
where either $\xi_ = 1$ or $\xi = -1$, according to whether the $i-th$ unit is in the cooperation or in the defection state. Of course, due to the fact that, as earlier stated,  we are using a finite number of units, $N = 1024$, the mean field fluctuates around the vanishing mean value.
As we increase $K_1$ the intensity of fluctuations tends to increase. This is a finite-size effect discussed in detail in the applications of the DMM model \cite{14,15}. In the case where each node is coupled to all the other nodes, the DMM yields analytical results for the mean field that can be interpreted as the space coordinate of a particle in a non-linear over-damped potential under the action of a random fluctuation generated by the finite size effect. At criticality, the potential 
is quartic, exerts a weaker containment on diffusion and makes larger the width of its equilibrium distribution.   The mean field fluctuation around the origin has an inverse power law with index $\mu = 1.5$: a form of temporal complexity ensuring the maximal efficiency in the transport of information  from one network to another with the same complexity\cite{16}. This important effect has been proved also in the case of a neural network model \cite{17}, thereby suggesting 
that criticality-induced temporal complexity may be a condition of great importance for the sensitivity of a complex system to its environment. Temporal complexity must not be confused with critical slowing down \cite{15}. Both properties are manifestations of criticality and both properties are characterized by the emergence of an inverse power law, which makes the survival probability not integrable. However, critical slowing down is a property of the thermodynamic limit, implying that the number of units $N$ is virtually infinite, whereas temporal criticality is a finite size effect \cite{21}. Temporal complexity, in the ideal case where the inverse power law is not truncated has to be thought of as a form of perennial out of equilibrium condition, which  should force physicists to extend the linear response theories to the non-ergodic condition \cite{7}.

Note that the cooperator and the defector payoffs are determined by the states of the nearest neighbors. Thus, we have
\begin{equation} \label{cooperator}
\Pi_C = (1 + \lambda) \frac{M_C}{M} - s \frac{M_D}{M}
\end{equation}
and
\begin{equation} \label{defector}
\Pi_D = (1 + t) \frac{M_C}{M} - \lambda \frac{M_D}{M},
\end{equation}
where $M_C$ is the numbers of neighbors in the cooperative state and $M_D$ is the number of neighbors in the defector state. We remind the readers that in this Letter $M= 4$.  It is important to state that we evaluate also the financial benefit for the community by making an average over all possible pairs of interacting units, according to the prescription:
\begin{equation} \label{cc}
B_{ij} =  2,
\end{equation}
if both units of the pair $(i,j)$  are cooperators,
\begin{equation} \label{cd}
B_{i,j} = 1 + t  - s,
\end{equation}
if one unit of the pair $(i,j)$ is a cooperator and the other is a defector,
\begin{equation} \label{dd}
B_{i,j} = 0,
\end{equation}
if both units of the pair $(i,j)$  are defectors.

Notice that the choice between the cooperation and defection state is done with $\lambda \geq 0$, while the financial benefits for the society are evaluated, as shown by Eqs. (\ref{cc}), (\ref{cd}) and (\ref{dd}),
by setting $\lambda = 0$. This is so because $\lambda$, the psychological reward, is an incentive to cooperate that does not directly increase the social payoff, even if it has the eventual effect of increasing the society wealth by stimulating cooperation.    In conclusion, the societal benefit $\Pi$ is given by
\begin{equation}
\Pi = \sum_{(i,j)} B_{ij},
\end{equation}
denoting the sum over all possible pairs $(i,j)$. 

To define the \emph{selfishness model} we set $\lambda = 0 $ and we establish the interaction between the bottom and the top level by assigning a positive value to the coupling coefficient $K_2$. In this case, a cooperator is encouraged  to adopt the cooperator state  for a longer time
if the financial benefit associated the choice of the cooperator condition, $\Pi_C$, is larger than the financial benefit  $\Pi_D$, corresponding to selecting the defector state. 
The \emph{success model} is established by setting both $\lambda = 0$ and $K_2 = 0$. The influence of the top on the bottom level is established by randomly selecting a fraction $r$ of the total number $L$ to allow  each unit 
to adopt the
strategy of the most successful nearest neighbor. We note that small values of $r$ play the same role as small values of $\rho_K$ in the selfishness model.  The influence of morality on the selfishness model is studied 
by setting $\lambda > 0$, while keeping $\lambda = 0$ for the evaluation of the social benefit, as earlier stated.

\begin{figure}
\onefigure[width= 0.4 \textwidth]{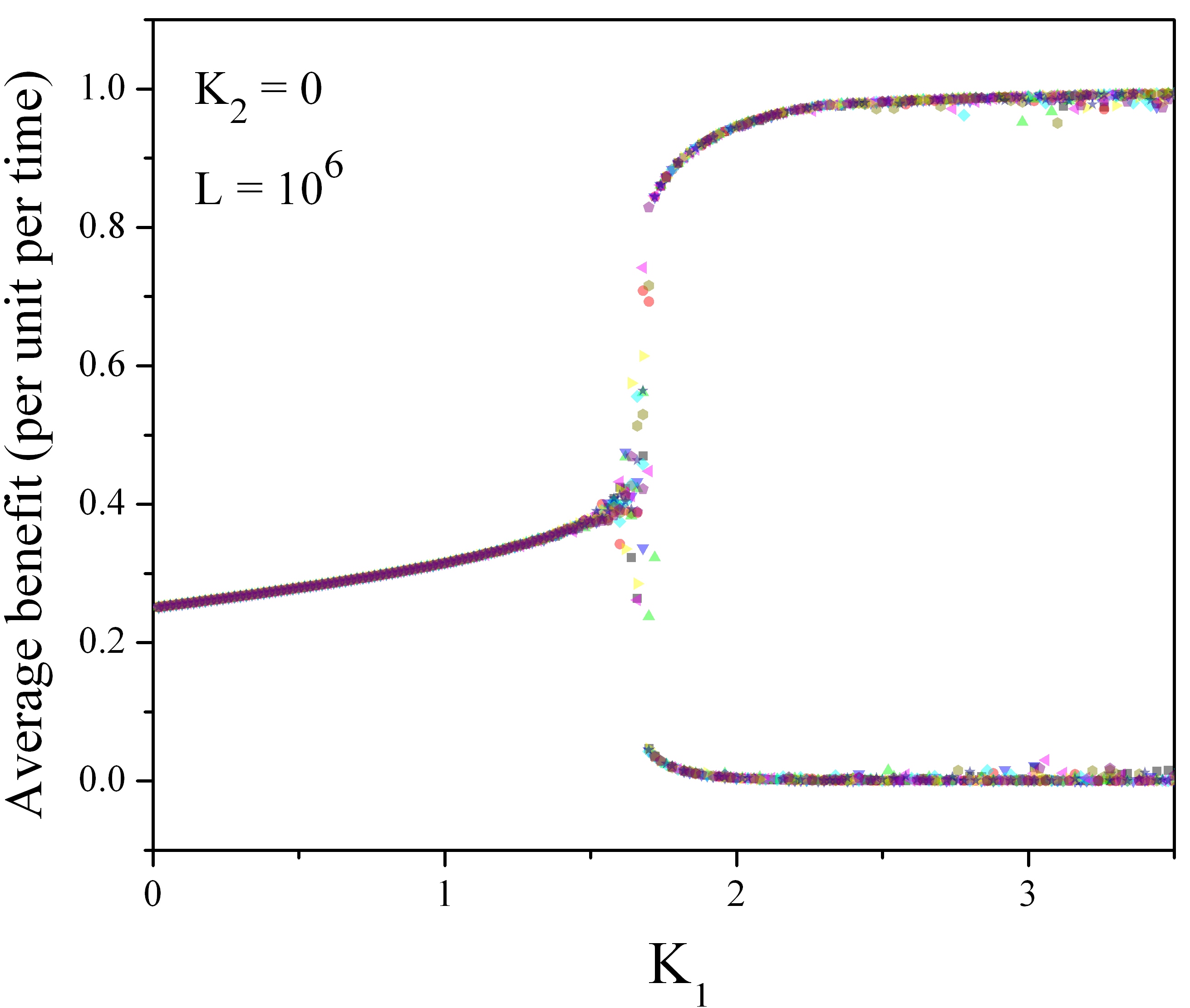}
\caption{Social benefit as a function of $K_1$ when $K_2 = 0$. We make ten realizations. We notice that the social benefit increases upon increase of $K_1$ in the whole subcritical region.}
\label{FIG2}
\end{figure}

\begin{figure}
\onefigure[width= 0.4 \textwidth]{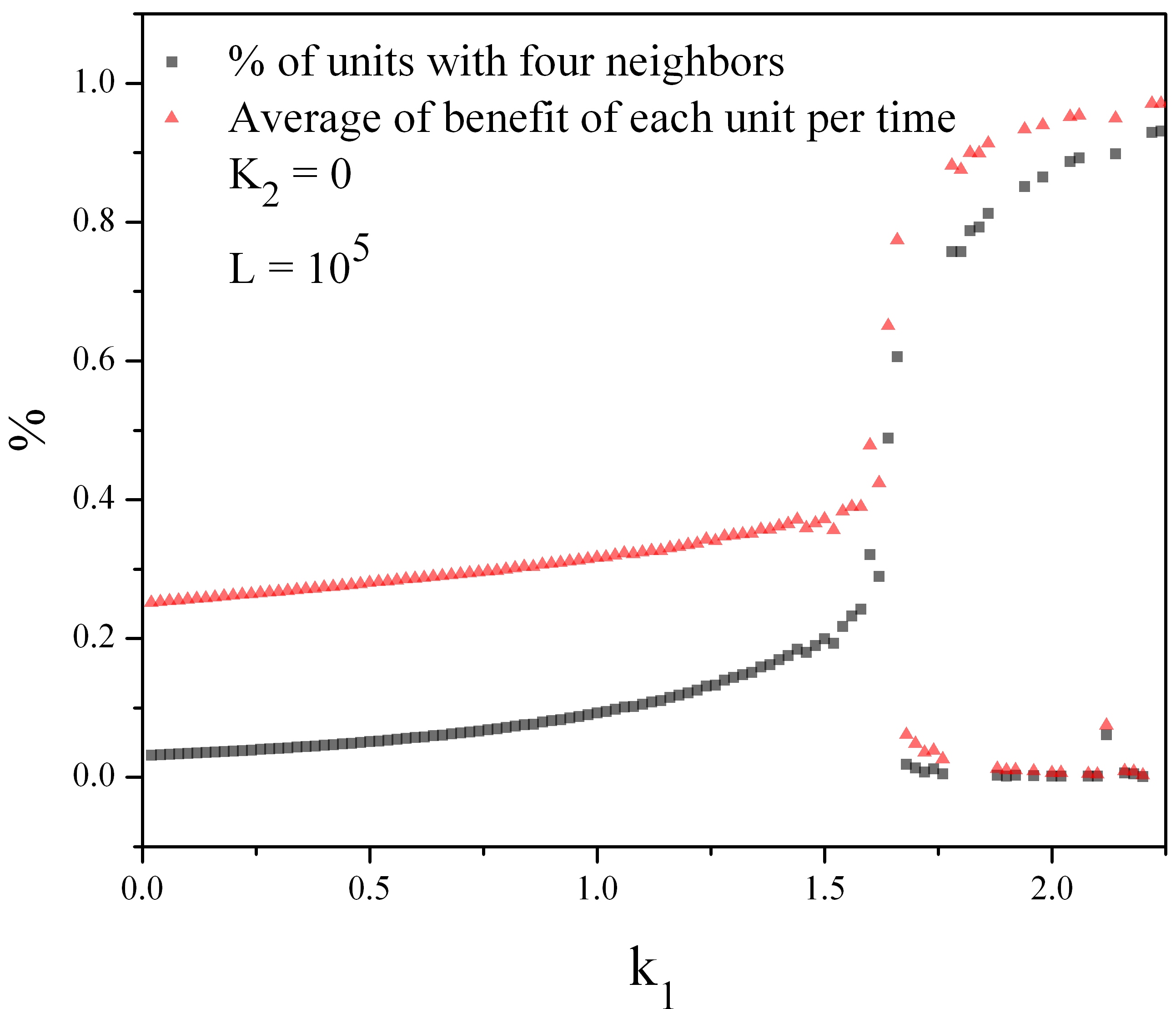}
\caption{ The  lines of the triangles  represents the social benefit. The line of black squares denotes the number of units with four neighbors in the same state. }
\label{FIG3}
\end{figure}

\section{Criticality-induced network reciprocity} 
In Fig. \ref{FIG2} we show the social benefit as a function of $K_1$ when $K_2 = 0$, namely, the strategy choice is only determined by imitation.  The social benefit in the supercritical regime is obviously maximal 
when all the units select the cooperation strategy and it vanishes when all the units select defection. Much more interesting is the social benefit for $K < K_{1c}$.  
We see that there is an increase of social benefit with increasing $K_1$. This is a consequence of the fact that imitation generates clusters of cooperators and clusters of defectors and the increasing social benefit is due to the increasing size of the clusters of cooperators. 

The results of Fig. \ref{FIG3} support our claim. In fact, we 
see that the social benefit increases in a way that is qualitatively very similar to the increase of the number of units 
that have four neighbors in the same state, either cooperation or defection state. A cluster of units belonging to the same state increases as a function of $K_1$ with a prescription
qualitatively similar to the increase of the number of units with four neighbors in the same state.

\begin{figure}
\onefigure[width= 0.4 \textwidth]{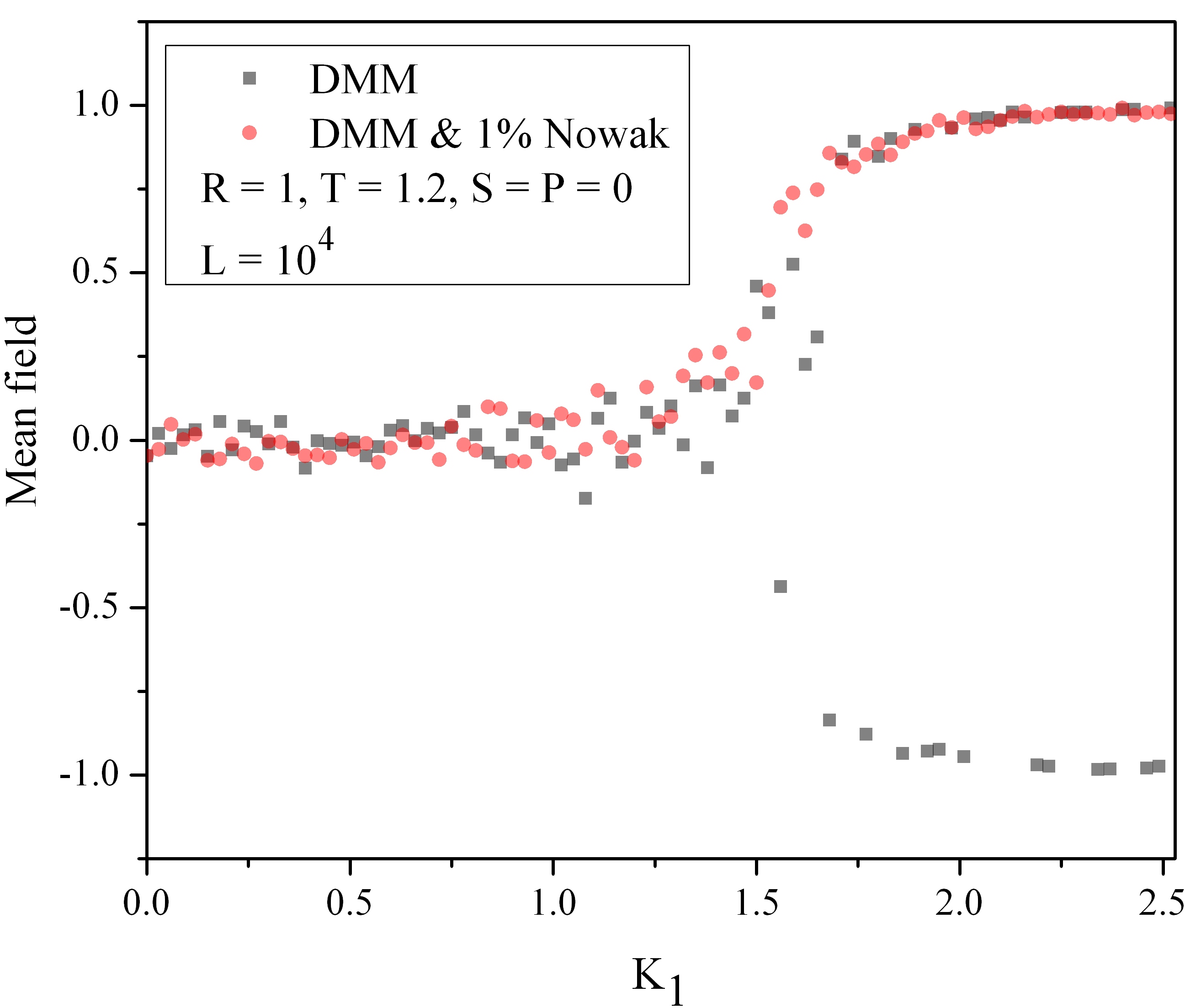}
\caption{The mean field of the imitation model as a function of $K_1$ randomly selecting $0.1 \%$ of $L$ steps to update the strategy of each unit by adopting the strategy of the most successful nearest neighbor. }
\label{NEW}
\end{figure}

We now adopt the \emph{success model}, namely, we perturb the imitation-induced strategy choice making the units pay some attention to the success of their nearest neighbors. We run the model for $L$ time steps, and we randomly select $1 \%$ of them to update their strategy adopting the  one of their most successful nearest neighbor. As earlier stated, 
we use a two-dimensional regular network, where each unit has
$4$ neighbors. The results illustrated by Fig. \ref{NEW} are impressive. The adoption of the strategy of the most successful nearest neighbor has a very modest effect in the subcritical region. At criticality, on the contrary, the effects of this choice become macroscopic and the imitation-induced phase transition, when the two branches, one with a majority of cooperator and one with a majority of defectors, are generated with equal probability, the success model selects the branch with a majority of cooperators. At very large values of $K_1$ this leads to the extinction of defectors. On the basis of the results done by our group on the DMM dynamics
we make the very plausible conjecture that this effect is independent of the topology of the adopted network. In fact,
moving from one topology to another has only the effect
of reducing the intensity of the effort necessary to get consensus,
the condition $K_1 = 1$ representing  the ideal topology requiring the weakest effort to get consensus \cite{8}.

\section{Morality stimulus on the selfishness model at criticality} 
As earlier mentioned the parameter  $\lambda$ is only a psychological benefit implying no direct financial benefit for society.  We interpret $\lambda$ as the strength of the influence that
the morality network has on the selfishness model.

\begin{figure}
\onefigure[width= 0.4 \textwidth]{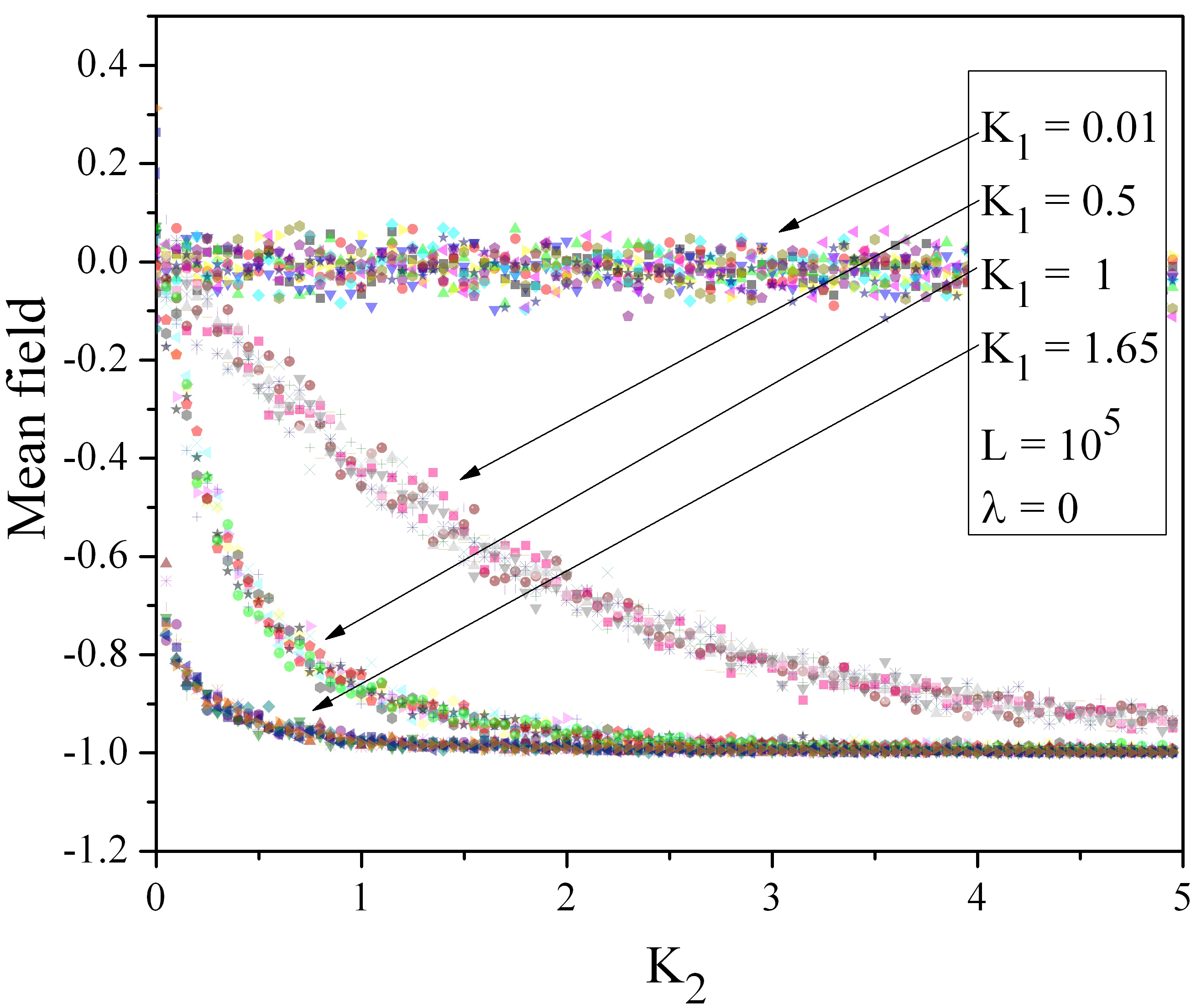}
\caption{  The mean field of the selfishness model as a function of $K_2$ for different values of $K_1$. We make ten realizations.}
\label{FIG5}
\end{figure}

Here we find the imitation-induced intelligence has a twofold effect. The collective intelligence is proved to lead to the abrupt extinction of cooperators, if $\lambda = 0$, and to the abrupt 
extinction of defectors when $\lambda = \lambda_c$. With the values of the parameters adopted in this Letter, $\lambda_c = 2.5$. 
 Fig. \ref{FIG5}, illustrating the effect of a strategy choice 
 when $\lambda = 0$,  shows that in the sub-critical regime a very large value of $\rho_K$ is required for the extinction of cooperation.  As we approach criticality, namely as the system becomes more and more intelligent, an even very weak interest for the personal financial benefit leads to the extinction of cooperators, which, in fact, is shown to occur for $\rho_K \approx 0.06$ when $K_1 = 1.6$. 

\begin{figure}
\onefigure[width= 0.4 \textwidth]{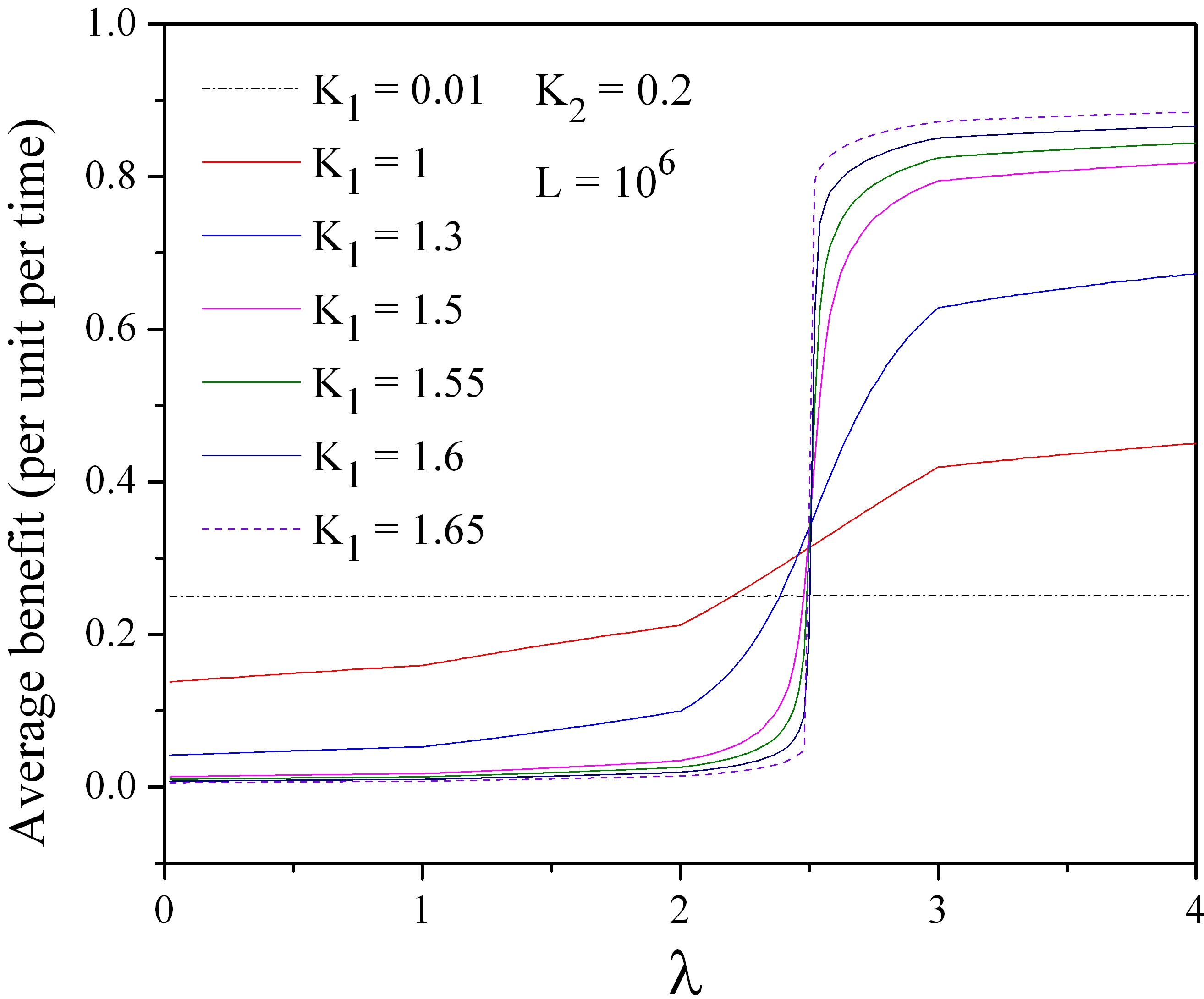}
\caption{ The personal benefit of the selfishness model as a function of $\lambda$. We make an average on ten realizations. }
\label{FIG6}
\end{figure}

The swarm intelligence of the system makes the model  very sensitive to the influence of morality, as it is clearly shown by Fig. \ref{FIG6}.
We see, in fact, that as consequence of the imitation-induced criticality, the average social benefit undergoes a kind of first-order transition at  $\lambda_c = 2.5$, with a jump from the lack of benefit to a very large value, when $K_1 = 1.65$ and $\rho_K \approx 0.12$.   This indicates that the criticality-induced intelligence wisely turns the 
psychological reward for the choice of cooperation into
a significant social benefit.

\begin{figure}
\onefigure[width= 0.4 \textwidth]{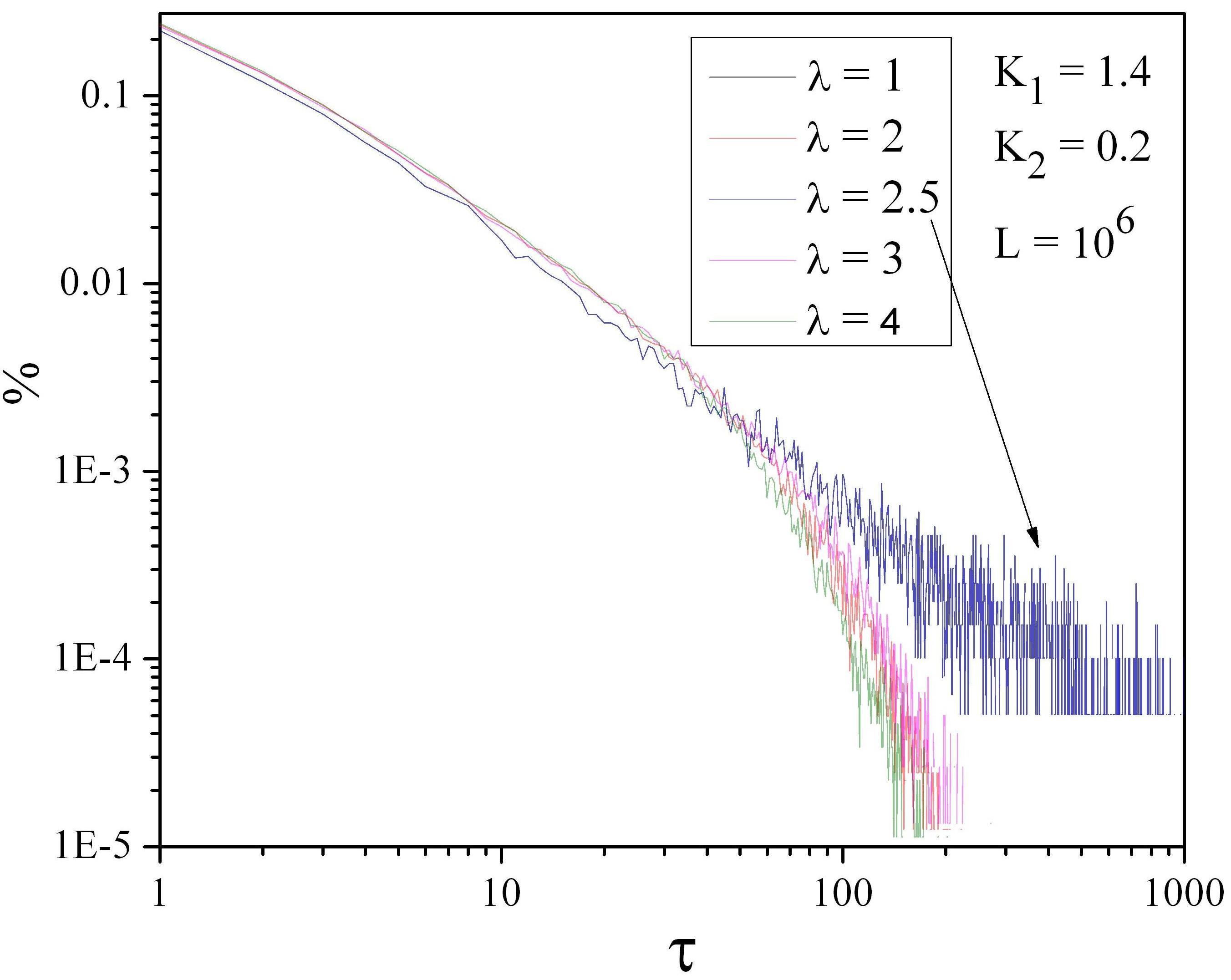}
\caption{ The cumulative probability of the time distances between two consecutive regressions to the origin. }
\label{FIG7}
\end{figure}

However, a value of $\lambda$ too large may have the same bad effects as a value of $\lambda$ too small. The corresponding cumulative
probability is illustrated by Fig. \ref{FIG7}, which shows that 
the distribution density of the time distances between two consecutive renewal events is an inverse power law with the power index $\mu = 1.5$ when $\lambda = \lambda_c = 2.5$ and it is an exponential function for both $\lambda < \lambda_c$ and $\lambda > \lambda_c$.  This is a clear sign that the collective intelligence generated by criticality \cite{14,15,16,17} is lost if the moral incentive to altruism $\lambda$ is either weak or excessive. 

\section{Concluding Remarks}
In conclusion we have proved that imitation-induced criticality 
has the effect of enhancing the phenomenon of network reciprocity. The adoption of the strategy of the most successful nearest neighbor not only protects the cooperator from extinction, as in the pioneer work of Nowak and May \cite{4}, but, at criticality, it annihilates the branch with the majority of defectors.  If we adopt the selfishness model for the choice of strategy,  the imitation-induced criticality has the effect of favoring the extinction of cooperators. Under the influence of morality stimulus, however, imitation-induced criticality has the opposite effect of leading to the extinction of defectors. However, the temporal complexity of the system is lost for both $\lambda < \lambda_c$ and $\lambda > \lambda_c$, 
indicating, in accordance with Ahmed \cite{20}, that a condition of super-asabiyya is detrimental for human society as the lack of asabiyya.

\acknowledgments
The authors warmly  thank ARO  for support through Grants No. W911NF-11-1-0478.

\end{document}